\documentclass[10pt]{article}
\usepackage[margin=2cm]{geometry}
\usepackage{lipsum}
\usepackage{amsfonts}
\usepackage{amsmath}
\usepackage{easy-todo}
\usepackage{wrapfig}
\usepackage{tcolorbox}
\usepackage{url}
\usepackage{caption}
\usepackage[labelformat=simple]{subcaption}
\usepackage{tabularray}

\usepackage{hyperref}
\usepackage{cleveref}
\usepackage{cclicenses}

\usepackage{graphicx} % Required for inserting images

\title{\textbf{Transcribing Educational Videos Using Whisper} \\
\Large{\sl A preliminary study on using AI for transcribing educational videos}}
\author{Ashwin Rao \\
University of Helsinki \\
}
\date{}

\begin{document}
\maketitle

\vspace{-2em}

\begin{abstract}
Videos are increasingly being used for e-learning, and transcripts are vital to enhance the learning experience. 
The costs and delays of generating transcripts can be alleviated by automatic speech recognition (ASR) systems. 
In this article, we quantify the transcripts generated by whisper for 25 educational videos and identify some open avenues of research when leveraging ASR for transcribing educational videos. 
\end{abstract}

\section{Introduction}

During the last decade, we have witnessed an increase in the volume of video content that is disseminated over the Internet. 
The pandemic further exacerbated this trend as people started to consume a wide category of videos from their houses~\cite{feldman:2020:covidinternet}. 
Along with lectures, we have also witnessed a rise in the conferences and talks that are being recorded and uploaded online on streaming sites.
These videos augment the material taught in the classrooms and are increasingly being leveraged for educational purposes~\cite{seaton2014characterizing}. 

Educational videos, like entertainment videos, are consumed in a combination of personal devices such as laptops, tablets, smartphones, and studies.
The capabilities of the audio systems on these devices vary significantly, and a given audio file may sound different on each of these devices~\cite{vox:subtitles}.
Words in an audio segment recorded by amateurs may sound clear and comprehensible on one device, and the same audio segment may be unintelligible on another device. 
Furthermore, the educational videos might include the voices of people from a wide range of ethnicities, and the speakers might also not be native speakers of the language in which they are speaking. 
Clearly, the audio quality of educational videos is vital, and addressing acoustic issues can result in drastic improvement in the quality of the material~\cite{richardson1998improving}.
However, the video and audio quality of educational videos might not be optimal for all devices because they may not be professionally created, edited, and processed.

\begin{wrapfigure}{l}{0.35\textwidth}
\centering
\begin{tcolorbox}
\small
\begin{verbatim}
WEBVTT
Kind: captions
Language: en

00:00:00.040 --> 00:00:02.460
The following content is
provided under a Creative

00:00:02.460 --> 00:00:03.870
Commons license.
\end{verbatim}
\end{tcolorbox}
\caption{\textbf{Example Closed Caption}. \textit{The metadata (the file format and language) is followed by the time stamps during which the text can be shown.}}
\label{fig:my_label}
\end{wrapfigure}

Audio transcripts and subcaptions help alleviate the issues in the audio quality and enable the viewers to receive a correct interpretation of the content.
For instance, Gernsbacher has shown that captions are particularly beneficial for persons watching videos in their non-native language~\cite{gernsbacher2015video}.
Although generating transcripts has been non-trivial, recent advances in speech-to-text generation have shown promising results in transcribing audio content. 
In the context of videos, transcripts are different from subtitles: transcripts typically refer to a textual copy of the words someone has said in the video, while subtitles refer to the textual versions of the dialogues in the video~\cite{wiki:subtitles}. 
Subtitles can either be open or closed: open subtitles are embedded in the video frames, while closed subtitles are stored separately and can be overlayed over the video frames or can be displayed on a second screen.
A variant of closed subtitles is closed captions which contain an additional description of the audio-video content being shown, such as the sound made by animals, etc.
At times, a transcript can also include additional description; examples include \textsl{laughter by students}, \textsl{audience clapping}, etc.  
A key difference between a transcript and the subtitles is that a transcript does not contain the time stamp at which the words in the transcript were said.

In this article, we do a preliminary evaluation of the quality of transcripts generated by whisper~\cite{radford:2022:whisper}.
We focus on the speech-to-text translation, and not on the time stamp at which the word was spoken.
Although there is a wide range of tools and models for generating transcripts, we focus our attention on whisper. 
Our goal is to get an understanding of using whisper for academic videos and identify open avenues of research in the area of leveraging ASR for transcribing academic videos. 

% \begin{figure}
%     \centering
%     \includegraphics[width=0.5\textwidth]{figures/plot-bitratecdf.pdf}
%     \caption{Caption}
%     \label{fig:my_label}
% \end{figure}

\section{Methodology}

\paragraph{Tools used and data processing pipeline.}
For our analysis, we first collect a set of 25 YouTube videos that have closed captions that are not automatically generated; YouTube shows if the captions are auto-generated or provided by the content creator.
For each video, we use \texttt{yt-dlp} to download the best audio files corresponding to the video and the available captions (as transcripts).
The downloaded captions are the baseline for our evaluation.
We do this because YouTube keeps multiple versions of the same video, and dynamically adapts to the optimal audio/video quality depending on the network connectivity.
We then use whisper~\cite{radford:2022:whisper} to generate the transcripts, and run it in our cluster powered by NVidia V100 GPUs~\cite{it4science}.
The generated transcripts are then compared with our baseline transcripts downloaded from YouTube using \texttt{jiwer}.
We summarize the tools used in \Cref{tab:tools}. 

\begin{wraptable}{l}{0.55\textwidth}
    \centering
    \begin{tabular}{l|l|l}
     \textbf{Tool}    &  \textbf{Version} & \textbf{Usage}\\
     \hline
     whisper & 20230314 & Speech to text conversion. \\
     jiwer & 3.0.1 & Compare the text in two files.\\
     yt-dlp & 2023.03.04 & Download audio files and transcripts. \\
     opusinfo & 0.1.10 & Extract metadata from audio files.\\
    \end{tabular}
    \caption{Software Tools}
    \label{tab:tools}
\end{wraptable}

\paragraph{Automatic Transcript Generation (Speech to Text).}
In this article, we restrict ourselves to whisper~\cite{radford:2022:whisper}.
Whisper offers multiple models which can be used to process the transcribe the audio files, and in our evaluation we restrict ourselves to the following five models (number of parameters in parenthesis) of which large-v2 is a multi-lingual model: base.en (74~M), tiny.en (39~M), small.en (244~M), medium.en (769~M), and large-v2 (1550~M). 
We acknowledge that there is a wide range of open-source tools and models including Kaldi~\cite{povey2011kaldi}, Flashlight~\cite{kahn2022flashlight}, and Paddlespeech~\cite{zhang2022paddlespeech}.
We plan to analyze the efficiency of these tools in our subsequent works.

\paragraph{Metrics for evaluating transcript quality.}
The Word Error Rate (WER) is a commonly used metric for comparing texts~\cite{morris:2004:wer} and it is computed as 
$WER = \dfrac{S+D+I}{N = H+S+D}$ where $H$ is the number of hits (correct words), $S$ is the number of substitutions, $D$ is the number of deletions, and $I$ is the number of insertions, and $N$ denotes the number of words in the reference (baseline) against which the hypothesis (results of the transcribing tool) is being evaluated. 
In contrast, the Match Error Rate (MER) is the probability of an incorrect match~\cite{morris:2004:wer}, and is given by $MER = \dfrac{S+D+I}{H+S+D+I}$.
The Word Information Lost (WIL) is an approximation for the Relative Information Lost (RIL) which is computed using the hits, substitutions, insertions, and deletions~\cite{morris:2004:wer}; the RIL measures the statistical dependence between the reference and the hypothesis and is calculated using the Shannon entropy.
Our goal is not to compare the metrics, and instead we rely on the WER, MER, and WIL to evaluate the performance of the transcription. 
We use jiwer to compute the WER, MER, and WIL. 
It is known that jiwer can end up computing a higher WER without normalizing the text~\cite{radford:2022:whisper}, and the WER depends on the normalization technique used. 
For this preliminary analysis we avoid using any custom normalizations, and we plan to explore the impact of normalization in a subsequent study. 

\begin{wrapfigure}{r}{0.5\textwidth}
    \centering
    \includegraphics[width=0.5\textwidth]{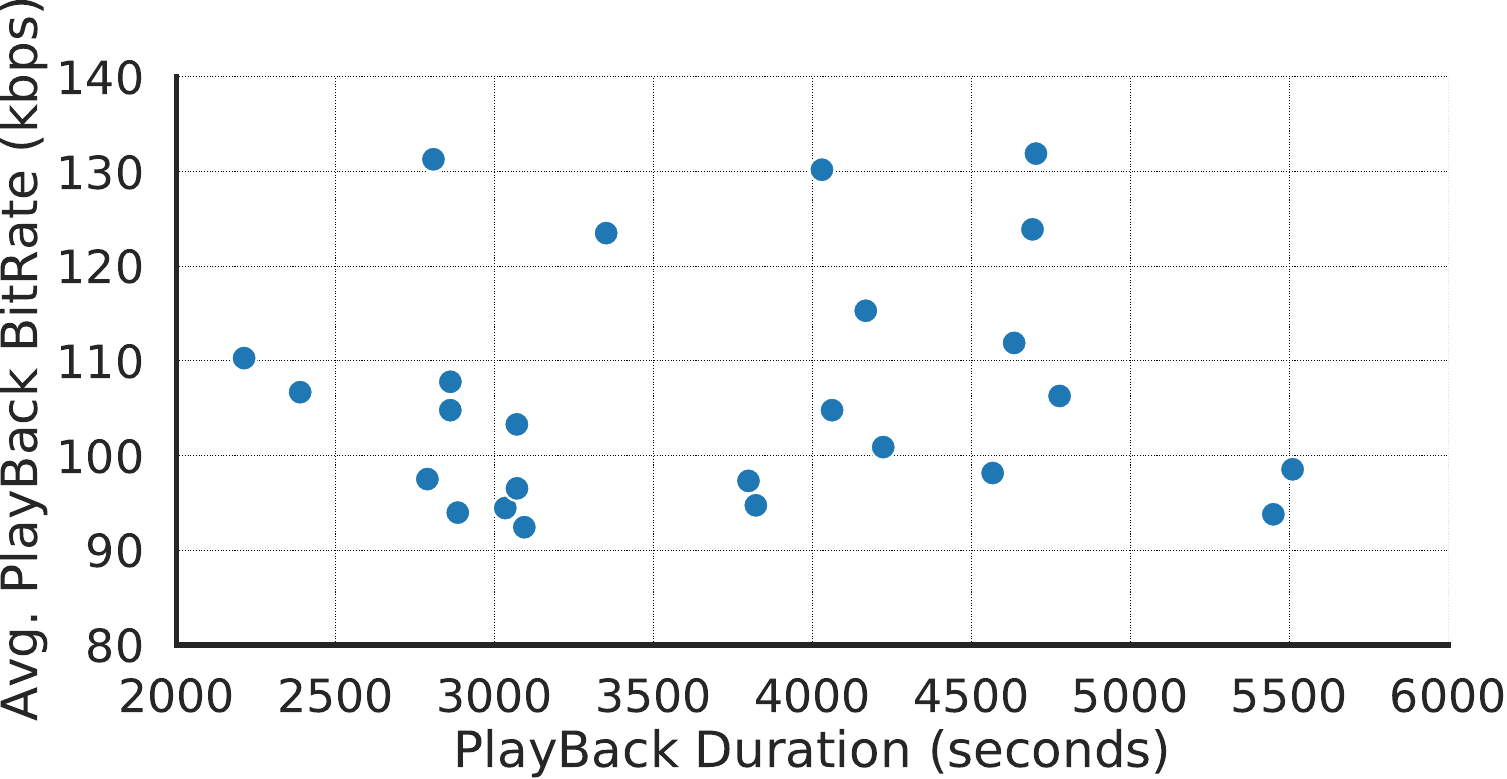}
    \caption{\textbf{Average Bitrate of the Audio Files.}}
    \label{fig:plt-bitrate-duration}
\end{wrapfigure}
\paragraph{Dataset Description.}
Of the 25 YouTube videos, 15 were from lectures on MIT OCW.
The remaining 10 included 5 talks at Google, one talk at MIT OCW, and four Turing Award lectures.\footnote{\textbf{Availability:} 
The details of these videos are available with our code and datasets at:
\url{https://version.helsinki.fi/transcribe-educational-videos/preliminary-study-dai2023/}}.
In \Cref{fig:plt-bitrate-duration}, we present the playback duration (size in seconds) of each of the videos and the average bitrate of the audio file. 
The quality of the audio file is important because it can affect the quality of the transcripts being generated, and we observe that the audio files downloaded have an average bit rate of at least 92 kbps. 
Note that the audio file was encoded in \texttt{opus} audio format which supports variable bitrate and is optimized for speech~\cite{opus}.
We also observe that the audio files were sampled at 48~kHz.
Whisper internally converts the audio file to 16~kHz, and we believe that the audio files in our dataset have a sufficiently higher frequency from which audio segments can be sampled at 16~kHz.

\section{Evaluation}

\begin{figure} [h] 
    \centering
    \begin{subfigure}[b]{0.48\textwidth}
    \includegraphics[width=\textwidth]{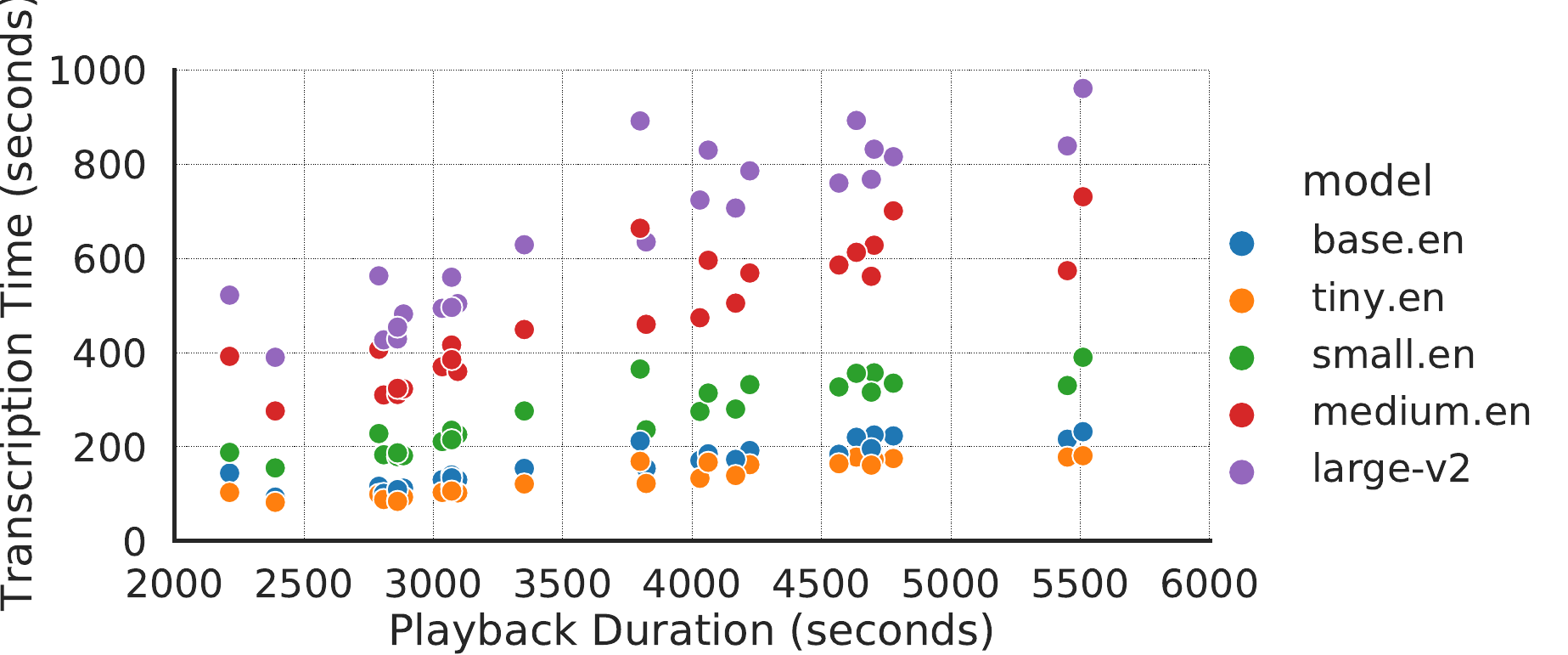}
    \caption{Time to transcribe a video for a given playback duration. }
    \label{fig:duration-proctime}
    \end{subfigure}
    \hfill
    \begin{subfigure}[b]{0.48\textwidth}
    \centering
    \includegraphics[width=\textwidth]{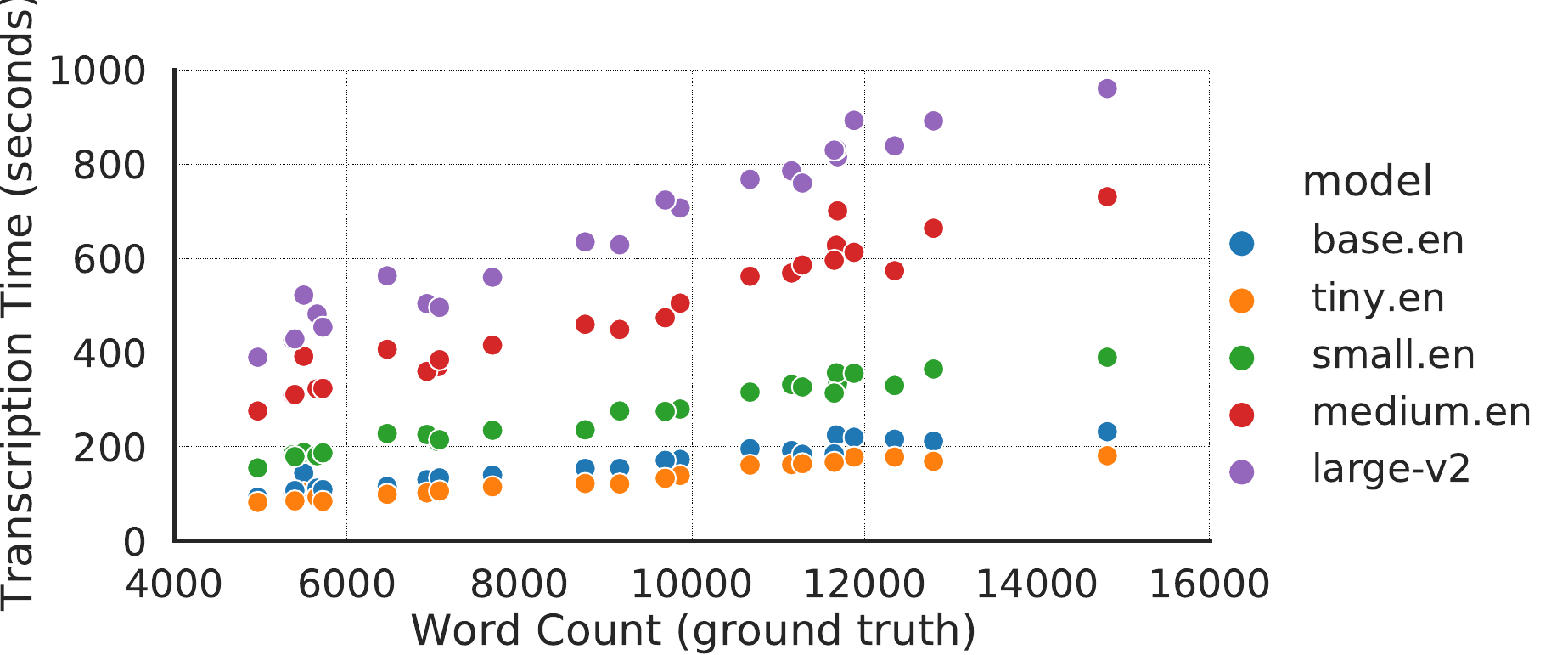}
    \caption{Time to transcribe a video for a given word count. }
    \label{fig:wc-proctime}
    \end{subfigure}
\caption{\textbf{Transcription Time.} \textit{The transcription time, i.e., the time to generate transcripts, increases linearly with the playback duration and word count. The larger models require more time than their smaller counterparts.}}   
\label{fig:proctime}
\end{figure}

%We first present the time required by whisper to generate the transcripts for the various models. 
In \Cref{fig:proctime}, we present the time required to transcribe a video for a given playback time (see \Cref{fig:duration-proctime}), and also for a given word count in our baseline transcripts (see \Cref{fig:wc-proctime}). 
We observe that the time to transcribe increases linearly with the playback duration and word count, and the larger models require more time.
We present these results to give a ballpark on what to expect, and we are aware that these times are heavily biased to the audio content, and the computational capabilities in our cluster. 

\begin{wrapfigure}{r}{0.45\textwidth}
    \centering
    \includegraphics[width=0.45\textwidth]{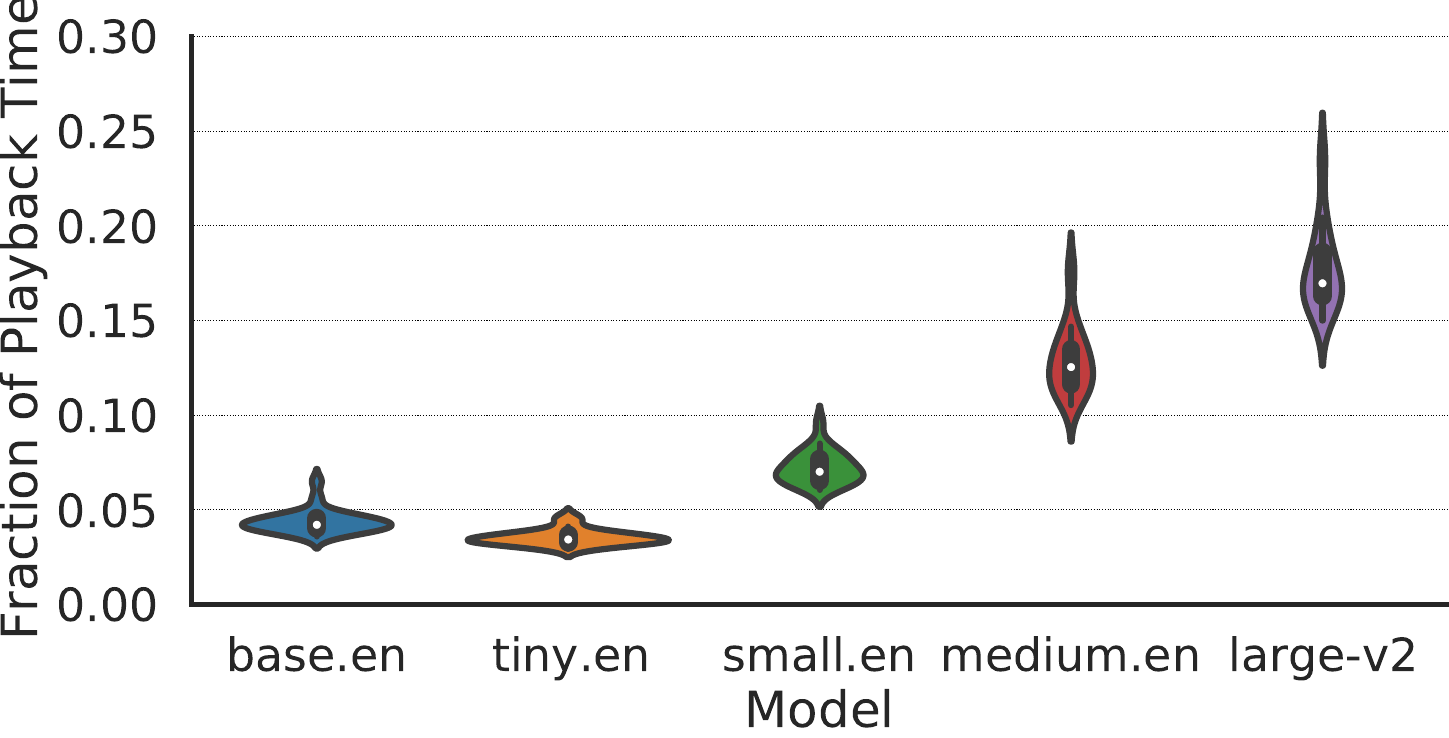}
    \caption{\textbf{Relative transcription time.} \textsl{If the playback time is 50~s and it takes 10~s to generate the transcript then the fraction of playback time is 10/50 = 0.2, i.e., generating a transcript required 20\% of the playback time. (Range = min, max)}}
    \label{fig:captionspeedup}
\end{wrapfigure}
In \Cref{fig:captionspeedup}, we plot the fraction of the playback time that a given model took to transcribe the video.
We observe that even the large-v2 model was able to complete the transcription process in less than 25\% of the time required to playback the video. 
For the videos in our dataset, and while running whisper on our servers, we observe that the base, tiny, and small models took less than 10\% of the playback time to transcribe the video, and the larger models took less than 25\% of the playback time. 
A typical human transcriber would require at least the playback time to listen to the whole audio.
In \Cref{fig:example-transcript}, we present a snippet of the transcripts generated using Whisper.
In this snippet, the speaker asks the audience member to repeat what they said because of audio issues. 
We see that the original transcript marks the conversation as inaudible while the whisper tries to guess what is said, and the results vary with the model size. 
Clearly, this speed-up when using smaller models is meaningless if the quality of the transcription is poor.

\begin{table}[h]
%\begin{tcolorbox}
\begin{footnotesize}
\begin{tblr}{|p{\textwidth}|}
\hline
\underline{Reference:} \textcolor{darkgray}{I'm ---- [inaudible] student from Duke.  I'm working on domain specific architecture.}\\
\hline[dotted]
\underline{base.en:}
\textcolor{blue}{I'm ------- ----, I'm a PhD student from Duke I'm working on Thomas versus architecture There's this thing called swallowing the microphones, so you have to actually Say it again, what you do? So I'm ------ ----, a PhD student from Duke I'm working on Thomas versus architecture}\\
\hline[dotted]
\underline{tiny.en:}
\textcolor{orange}{I'm ---- --- ----, I'm a PhD student from Duke, I'm working on So I'm ---- --- ----, a PhD student from Duke, I'm working on Thomas's certificate architecture}\\
\hline[dotted]
\underline{small.en:}
\textcolor{teal}{I'm --- -------. I'm a PhD student from Duke. I'm working on There's this thing called swallowing the microphone so you have to actually yeah, so say it again what you do So I'm --- ------- a PhD student from Duke. I'm working on a domain specific architecture}\\
\hline[dotted]
\underline{medium.en:}
\textcolor{purple}{I'm ------- ----. I'm a PhD student from Duke. I'm working on domain-specific architecture. There's this thing called swallowing the microphones. You have to actually... Say it again. What did you do? I'm ------- ----, a PhD student from Duke. I'm working on domain-specific architecture.}\\
\hline[dotted]
\underline{large-v2:}
\textcolor{violet}{I'm ---- --- ---- I'm a PhD student from Duke I'm working on domain specific architecture there's this thing called swallowing the microphone so you have to
actually yeah so say it again what you do so I'm ---- --- ---- a PhD student from Duke I'm working on domain specific architecture.}\\
\hline
\end{tblr}
\end{footnotesize}
%\end{tcolorbox}
\caption{\textbf{Example transcript with high WER.} \textit{The above transcripts are for a segment at time offset 1h:02m:58s of the the following video} {https://www.youtube.com/watch?v=3LVeEjsn8Ts\#t=62m58s.}}
\label{fig:example-transcript}
\vspace{-1em}
\end{table}

\begin{wrapfigure}{l}{0.5\textwidth}
    \centering
    \includegraphics[width=0.5\textwidth]{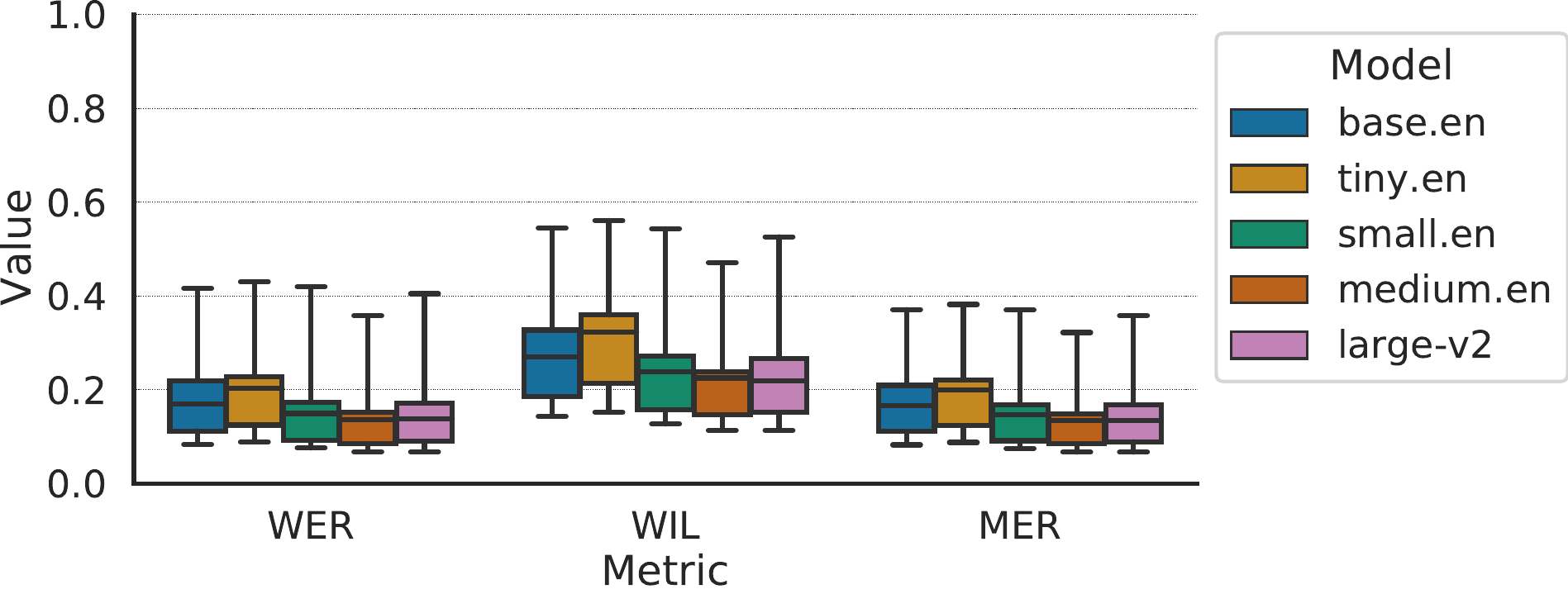}
    \caption{\textbf{Transcript quality}. \textit{The error bars represent the min and max across the files in the dataset.}}
    \label{fig:jiwer}
    \vspace{-1em}
\end{wrapfigure}
In \Cref{fig:jiwer}, we present the WER, MER, and WIL when using the various models.
Across all the metrics, we observe that the WER, MER, and WIL decreases as the number of parameters in the models increases. 
An exception is for the large-v2 model. 
We believe that this is primarily due to the lack of using a normalizer~\cite{radford:2022:whisper}, and the audio segments that were marked \texttt{inaudible} in the original transcripts. 
As shown in \Cref{fig:example-transcript}, whisper transcribes the conversation marked \texttt{inaudible} by the human transcriber, and the volume of text generated (sans punctuations) by the large-v2 model is larger than the other models thus resulting in a higher error rate.  

\begin{wrapfigure}{r}{0.6\textwidth}
    \centering
    \includegraphics[width=0.6\textwidth]{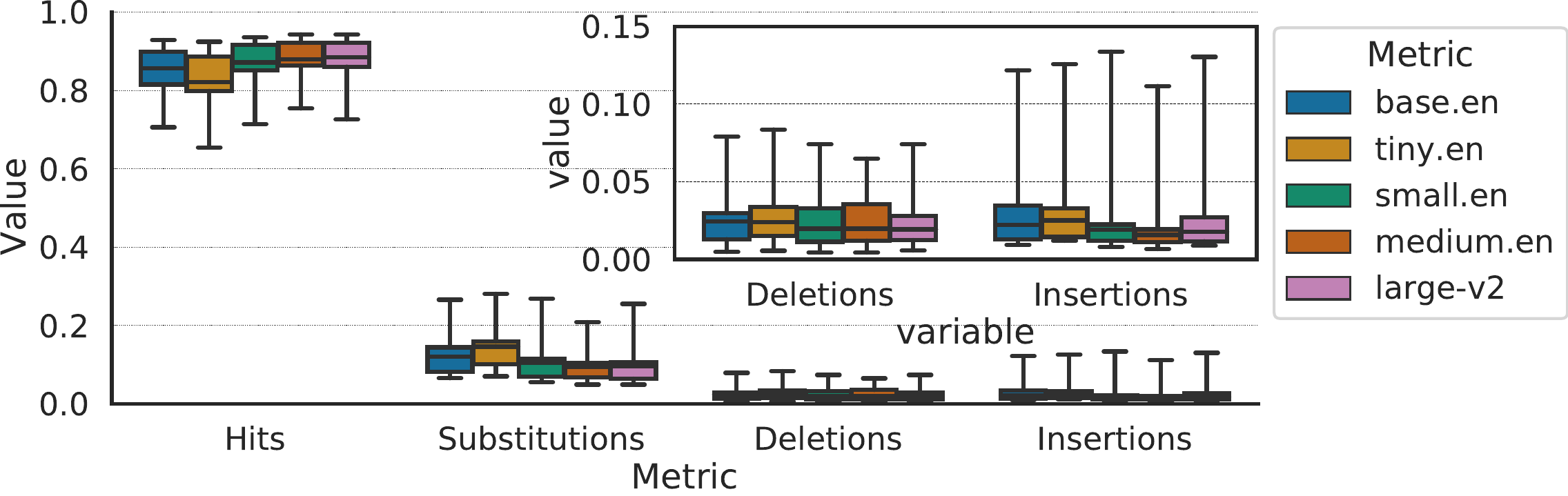}
    \caption{\textbf{Fraction of Hits, Substitutions, Deletions, and Insertions.} \textit{Error bars represent the min and max across files in our dataset. The cutout zooms into the Deletions and Insertions.}}
    \label{fig:hids}
    \vspace{-1em}
\end{wrapfigure}

Along with the example provided in \Cref{fig:example-transcript}, we also observe a high WER, a high WIL, and a high MER for other videos, as highlighted by the error bars in \Cref{fig:jiwer}.
To better understand this behavior, we present the fraction of hits, substitutions, deletions, and insertions in \Cref{fig:hids}.
Across all models, we observe that the hits are above 80\% for the majority of videos, and the fraction of hits increases with the number of parameters. 
However, for some videos, such as the one in \Cref{fig:example-transcript}, we observe a large number of substitutions, insertions, and deletions. 

One reason for the high error rates is that whisper does not provide \texttt{inaudible} as output and tries to extract the text even from the audio which a human transcriber might mark as \texttt{inaudible}. 
This is further exacerbated by not leveraging the context. 
For instance, in the example shown in \Cref{fig:example-transcript} the conversation was about domain-specific architecture, and the question being asked was on the same topic, and yet some of the models wrongly predicted the outcome to be \textit{Thomas version architecture} or \textit{Thomas's certificate architecture}.
These predictions are \textsl{bullshit}\footnote{We apologize for the use of profanity, and we rely on the following quote by Harry Frankfurt~\cite{frankfurt2005bullshit} for describing the term \textit{bullshit}: ``\textit{it is impossible for someone to lie unless he thinks he knows the truth. Producing bullshit requires no such conviction}.''} because they (and the underlying models) are indifferent to truth.
Furthermore, although only two substitutions are needed to replace \textit{thomas certificate architecture} to \textit{domain specific architecture}, incorrect predictions like these diminish the usefulness of the generated transcripts.
We believe that marking the audio segments as either \texttt{inaudible} or its equivalent that indicates a low confidence in the transcription result would be more beneficial in such scenarios.
This is achievable by tweaking some thresholds in whisper's configurations, and we plan to explore their impact in subsequent works. 

\section{Concluding Remarks and Avenues for Future Work}

We performed a preliminary analysis on the transcription capabilities of Whisper, however we cannot draw any strong conclusions: \textit{our dataset is heavily biased to the videos picked by the author, and the results are only for the models of one tool, whisper}.
However, we gained some insights such as the importance of marking audio segments as \texttt{inaudible}, and how \texttt{inaudible} audio segments affect the quality of transcripts generated by ASR systems.

Some avenues for future work in this area include: a) metrics that account for the semantic information such as the importance of each word, and evaluate the quality of transcripts in end-user studies; b) comparing the transcription results from different models; c) evaluating transcription capabilities for languages other than English, and also for non-native speakers for these languages; d) quantifying the impact of multiple speakers from different ethical backgrounds in the same video/audio; e) approaches to identify the context of the lecture/talk, and leveraging it for better transcriptions; f) quantifying the costs for generating transcripts for different accelerators, and identifying effectiveness of accelerators for transcript generation on end-user devices; and g) quantifying the quality of subtitles including the timestamp of the words and descriptions of the sounds that are generated by the ASR system. 

% Approaches to reduce the biases in the dataset.
% Approaches to reduce the biases in the tools used.

\paragraph{Acknowledgement.}
The authors wish to thank the Finnish Computing Competence Infrastructure (FCCI) for supporting this project with computational and data storage resources

%\section*{Availability}

\clearpage

\bibliographystyle{unsrt}
\bibliography{references.bib}

%\hline
\vfill

This work is licensed under a \href{https://creativecommons.org/licenses/by-sa/4.0/}{Creative Commons Attribution-ShareAlike 4.0 International License.}
\bysa

\end{document}